\documentclass{sf2a-conf2015}
\usepackage{graphicx}
\usepackage{hyperref}
\usepackage[]{natbib}  
\usepackage{epstopdf}

\def\BibTeX{{\rm B\kern-.05em{\sc i\kern-.025em b}\kern-.08em
    T\kern-.1667em\lower.7ex\hbox{E}\kern-.125emX}}
\bibpunct{(}{)}{;}{a}{}{,}  

\begin{document}

\TitreGlobal{SF2A 2015}


\title{The magnetic field of the hot spectroscopic binary HD\,5550}

\runningtitle{The magnetic field of HD\,5550}

\author{C. Neiner}\address{LESIA, Observatoire de Paris, PSL Research
University, CNRS, Sorbonne Universit\'es, UPMC Univ. Paris 06, Univ. Paris
Diderot, Sorbonne Paris Cit\'e, 5 place Jules Janssen, 92195 Meudon, France}

\author{E. Alecian$^{1,}$}\address{Univ. Grenoble Alpes, CNRS, IPAG, F-38000
Grenoble, France}

\author{the BinaMIcS collaboration}




\setcounter{page}{237}


\maketitle


\begin{abstract}
HD\,5550 is a spectroscopic binary composed of two A stars observed with Narval
at TBL in the frame of the BinaMIcS (Binarity and Magnetic Interactions in
various classes of Stars) Large Program. One component of the system is found to
be an Ap star with a surprisingly weak dipolar field of $\sim$65 G. The
companion is an Am star for which no magnetic field is detected, with a
detection threshold on the dipolar field of $\sim$40 G. The system is tidally
locked, the primary component is synchronised with the orbit, but the system is
probably not completely circularised yet. This work is only the second detailed
study of magnetic fields in a hot short-period spectroscopic binary. More
systems are currently being observed with both Narval at TBL and ESPaDOnS at
CFHT within the BinaMIcS project, with the goal of understanding how magnetism
can impact binary evolution and vice versa. 
\end{abstract}



\section{Observations of HD\,5550}

HD\,5550 is a spectroscopic double-line (SB2) binary system composed of two
A-type components \citep{carrier2002}. HD\,5550 was previously reported to be an
Ap SrCrEu star \citep{renson1991}. \cite{carrier2002} also reported that the
secondary has chemical peculiarities, but they could not distinguish more
precisely the peculiar type of this component.

We observed HD\,5550 in the frame of the BinaMIcS (Binarity and Magnetic
Interactions in various classes of Stars) project, with the goal to understand
the interplay between magnetism and binarity (see Neiner et al., these
proceedings). Twenty-five high-resolution spectropolarimetric observations were
obtained with Narval at the Bernard Lyot Telescope (TBL, Pic du Midi, France)
and were used to check for the presence of a magnetic field in both components.

We first disentangled the spectra of the two components to be able to analyse
them separately. The binary orbit has a period $P_{\rm orb}$ = 6.82054 d
\citep{carrier2002} and is almost circularised with an eccentricity e=0.005. We
then used Zeeman and Atlas9 LTE models on the disentangled spectra to derive the
stellar parameters of both components: we confirmed that the primary component
is an Ap star and found that the secondary is an Am star, with overabundance of
the iron-peak elements, extreme overabundance of Ba, and underabundance of 
Ca. Finally, we applied the Least-Square Deconvolution (LSD) technique to
produce averaged Stokes I and V spectra of each component and we measured the
magnetic field in both stars.

\section{Magnetic results}

\subsection{Primary Ap star}

\begin{figure}[t!]
 \centering
 \includegraphics[width=0.7\textwidth,clip]{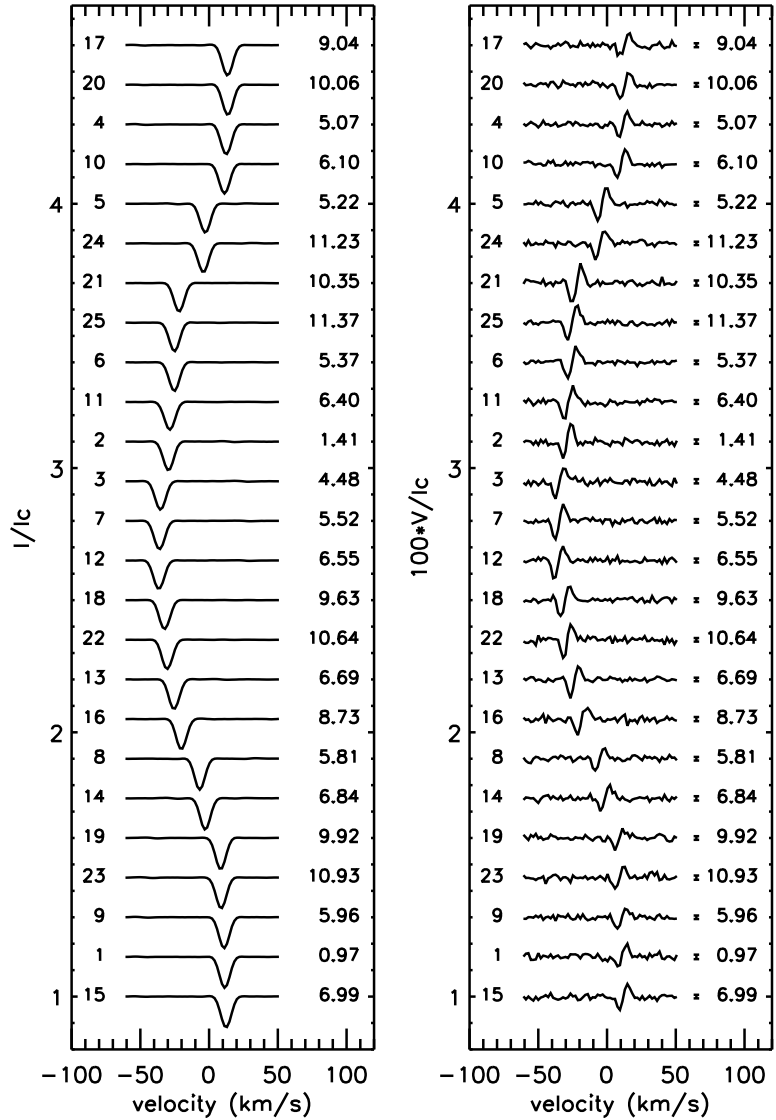}      
  \caption{LSD I/Ic (left) and V/Ic (right) profiles of the primary component of HD\,5550, ordered by increasing orbital phase. Taken from Alecian et al.,  submitted to A\&A.}
  \label{Neiner1:fig1}
\end{figure}

We found that the primary Ap star is magnetic with clear Zeeman signatures (see
Fig.~\ref{Neiner1:fig1}). The longitudinal field $B_l$ values are systematically
negative and vary from $-26$ to $-12$ G, with typical error bars of 4 G.

From the variations of the Stokes V profiles, and the corresponding $B_l$
values, we found that the field is modulated by the rotation period $P_{\rm rot}
\sim$ 6.84 d. This period is compatible with the orbital period $P_{\rm orb}
\sim$ 6.82 d, i.e. the rotation of the star is synchronised with the binary
orbit.

An oblique dipole model of the Zeeman signatures shows that the polar field
strength is only $B_{\rm pol}$ = 65 $\pm$ 20 G, with an inclination $i$ $\sim$
32$^\circ$ and an obliquity $\beta$ $\sim$ 156$^\circ$. This is the weakest
magnetic field known in an Ap star. Indeed, typical magnetic field strengths in
Ap/Bp stars are of the order of 1 kG, with a range between 300 G and 30 kG
\citep[e.g.][]{borra1980,  landstreet1992, bagnulo2006}. The dipolar field value
of HD\,5550 falls in the dichotomy desert proposed by \cite{auriere2007} between
strong and ultra-weak fields. 

\subsection{Seconday Am star}

We did not detect a magnetic field in the secondary Am star. The longitudinal
field values we measured by integrating the LSD I and V profiles are all
consistent with 0 G, with uncertainties of 3-4 G.

To determine the upper limit on the possible undetected magnetic field of the
secondary star, we first fitted the LSD I profiles with Gaussian profiles. We
then computed 1000 synthetic Stokes V profiles for various values of the polar
magnetic field $B_{\rm pol}$. Each of these models uses a random inclination
angle $i$, obliquity angle $\beta$, and rotational phase. We added a white
Gaussian noise to each modeled profile with a null average and a variance
corresponding to the signal-to-noise of the observed profile. We then computed
the detection probability of a magnetic field as a function of $B_{\rm pol}$ for
each observation, and combined them to obtain the detection probability function
for the full dataset. Above a 90\% detection probability, we consider that we
would have detected the field in our dataset. We therefore established that the
upper limit of the magnetic field of the secondary Am component, which could
have remained hidden in our observations, is $\sim$40 G.

Only a few Am stars are known to host a magnetic field so far and all of them
have ultra-weak fields, with longitudinal components of less than 1 G
\citep{petit2011, blazere2015}. If such an ultra-weak field were present in the
Am component of HD\,5550, it would have remained undetected in our observations.

\section{Conclusions}

Spectropolarimetric Narval observations of HD\,5550 showed that it is a binary
system composed of a weakly magnetic Ap star and an Am star found to be
non-magnetic with the achieved precision. With HD\,98088 \citep{folsom2013},
this is the second hot magnetic spectroscopic binary studied in details.
Studying more hot magnetic binaries, which is one of the goals of the BinaMIcS
project, will allow us to understand the interplay between magnetism and
binarity in hot systems.

\begin{acknowledgements}
We thank the ``Programme National de Physique Stellaire" (PNPS) of CNRS/INSU
(France) for their financial support to the BinaMIcS project.
\end{acknowledgements}

\bibliographystyle{aa}  
\bibliography{Neiner1} 

\end{document}